\documentclass[reprint,amsmath,amssymb,aps,prb,showpacs,ongbibliography]{revtex4-1}

\usepackage{graphicx}
\usepackage{epsfig}
\usepackage{dcolumn}
\usepackage{bm}
\usepackage{hyperref}
\hypersetup{backref=true,
 pdfnewwindow=true, colorlinks=true,
 linkcolor=blue, anchorcolor=blue,
 citecolor=blue, filecolor=blue,
 menucolor=blue, urlcolor=blue}

\begin{document}
\title{Wannier representation of $\mathbb{Z}_2$ topological insulators}
\author{Alexey A. Soluyanov}
\email{alexeys@physics.rutgers.edu}
\author{David Vanderbilt}
\email{dhv@physics.rutgers.edu}
\affiliation{Department of Physics and Astronomy, Rutgers University,
Piscataway, New Jersey 08854-0849, USA}
\date{\today}

\begin{abstract}
We consider the problem of constructing Wannier functions
for $\mathbb{Z}_2$ topological insulators in two dimensions.
It is well known that there is a topological obstruction to the
construction of Wannier functions for Chern insulators, but it
has been unclear whether this is also true for the $\mathbb{Z}_2$
case.  We consider the Kane-Mele tight-binding model, which
exhibits both normal ($\mathbb{Z}_2$-even) and topological
($\mathbb{Z}_2$-odd) phases as a function of the model parameters.
In the $\mathbb{Z}_2$-even phase, the usual projection-based
scheme can be used to build the Wannier representation.  In the
$\mathbb{Z}_2$-odd phase, we do find a topological obstruction,
but only if one insists on choosing a gauge that respects the
time-reversal symmetry, corresponding to Wannier functions that
come in time-reversal pairs.  If instead we are willing to violate
this gauge condition, a Wannier representation becomes possible.
We present an explicit construction of Wannier functions for the
$\mathbb{Z}_2$-odd phase of the Kane-Mele model via a modified
projection scheme followed by maximal localization, and confirm
that these Wannier functions correctly represent the electric
polarization and other electronic properties of the insulator.
\end{abstract}
\maketitle

\marginparwidth 2.7in
\marginparsep 0.5in
\def\dvm#1{\marginpar{\small DV: #1}}
\def\asm#1{\marginpar{\small AS: #1}}
\def\scr{\scriptsize}

\def\r{{\bf r}}
\def\R{{\bf R}}
\def\k{{\bf k}}
\def\G{{\bf G}}

\section{INTRODUCTION}

In the past several years there has been a surge of interest in topological
insulators.  These are materials that are gapped in the bulk, just like
ordinary insulators, but that cannot be adiabatically connected to ordinary
insulators without closing the gap or breaking some specified symmetries.
They also exhibit chiral metallic edge states that are topologically
protected from disorder.\cite{Schnyder-PRB08,Schnyder-AIP09,Kitaev-AIP09}
Topological insulators can be distinguished from normal ones based
on the manner in which the Bloch eigenfunctions are topologically
twisted in k-space.

Two types of topological insulators have received the most attention.
First, Thouless {\it et al.}\cite{Thouless-PRL82} pointed out long ago
that a two-dimensional (2D) insulator is characterized in general by a
topological integer known as the ``Chern number'' or ``TKNN index.''
A prospective insulator having a non-zero value of this integer would
be known as a ``Chern'' or ``quantum anomalous
Hall'' insulator.  The latter name
arises because such a crystal would exhibit a quantum Hall effect (QHE)
even in the absence of a macroscopic magnetic field, and would have chiral
edge states just like the ordinary field-induced QHE.  Haldane devised
an explicit tight-binding model realizing such a case.\cite{Haldane-PRL88}
Since the Hall conductance is odd under the time-reversal ($T$) operator,
Chern insulators can only be realized in systems with broken $T$ symmetry,
e.g., insulating ferromagnets.  Despite the fact that these possibilities
have been appreciated now for almost three decades, no known experimental
realizations of a Chern insulator are yet known.

Second, a great deal of interest has surrounded the recent discovery
of a different class of topological insulators known as $\mathbb{Z}_2$
insulators that realize the quantum spin Hall effect (QSH).\cite{Kane-PRL05-b} 
Subsequent theoretical\cite{Bernevig-Science06,
Zhang-NatPhys09,Fu-PRB07} and experimental\cite{Konig-Science07,
Chen-Science09,Hsieh-Nature08,Hsieh-Science09,Xia-NatPhys09}
work has succeeded in identifying several materials systems that realize
the case of a $\mathbb{Z}_2$ topological insulator.
Unlike the Chern index, which vanishes unless
$T$ is broken, the $\mathbb{Z}_2$ index
(which takes values of 0 and 1, or equivalently, ``even'' and ``odd'')
is only well defined when $T$ is conserved.  $\mathbb{Z}_2$ insulators
are thus non-magnetic, although a spin-orbit or similar interaction is
needed to mix the spins in a non-trivial way.  Because $T$ is preserved,
the occupied states at ${\bf k}$ and $-{\bf k}$ form Kramers pairs, and one can
associate a $\mathbb{Z}_2$ invariant with the way in which these Kramers
pairs are connected across the Brillouin zone.\cite{Moore-PRB07}  Since
the $\mathbb{Z}_2$ index cannot change along an adiabatic path that is
everywhere gapped and $T$-symmetric, a $\mathbb{Z}_2$-even (normal)
insulator cannot be connected to a $\mathbb{Z}_2$-odd (topological)
one by such a path.  In 2D there is a single $\mathbb{Z}_2$ invariant,
and $T$-invariant insulators are classified as ``even'' or ``odd,'' while
in 3D there are four $\mathbb{Z}_2$ invariants and the classification
is more complicated.\cite{Fu-PRL07}

Wannier functions (WFs) have proven to be a valuable tool when working
with semiconductors and insulators, providing a real-space description
that can be used to understand bonding, construct model 
Hamiltonians, and directly compute certain physical properties
such as the electric polarization.\cite{Marzari-PRB97, Vanderbilt-PRB93} 
Thus, it is desirable to understand the construction of the Wannier
representation for topological insulators so that this useful
set of techniques can be applied to these novel materials.
 
For Chern insulators it has been shown that a non-zero Chern number
presents a topological obstruction that prevents the construction of
exponentially localized
WFs.\cite{Thonhauser-PRB06,Thouless-JPC84} Conversely, a general proof
has been given that exponentially localized WFs should exist in any 2D
or 3D insulator having a vanishing Chern index.\cite{Brouder-PRL07}
In principle this applies to $\mathbb{Z}_2$-odd as well as
$\mathbb{Z}_2$-even $T$-invariant insulators, suggesting that a Wannier
representation should be possible in both cases.  However, it is unclear
whether the nontrivial topology of the $\mathbb{Z}_2$-odd case has any
effect on the Wannier representation. In particular, one may wonder
whether the procedure for obtaining WFs would be the same as for
ordinary insulators, and if not, how it should be modified in order to
get well localized WFs in the $\mathbb{Z}_2$-odd regime.

In this paper we address this question using the model of Kane and
Mele \cite{Kane-PRL05-b} as a paradigmatic system that exhibits both
$\mathbb{Z}_2$-odd and $\mathbb{Z}_2$-even phases.
We demonstrate that the usual
projection scheme used for constructing the Wannier representation is
still applicable to the $\mathbb{Z}_2$-odd insulators, but only for
gauge choices that do not allow WFs to come in time-reversal pairs. We
present an explicit projection procedure for constructing well-localized
WFs in the topologically non-trivial phase, and show that the WFs can be
made even more localized using the standard maximal-localization
procedure.\cite{Marzari-PRB97}  We also discuss the electric polarization
from both Berry-phase and Wannier points of view, showing the relations
between the viewpoints and confirming that both give identical results.

The paper is organized as follows. In Sec.~\ref{Sec.2} we define the
$\mathbb{Z}_2$ topological invariant in 2D and briefly discuss methods for
determining it numerically.  We review the model of Kane and Mele
in Sec.~\ref{Sec.3}, and describe its spectrum and phase diagram.
In Sec.~\ref{Sec.4} we present the projection scheme used to construct
WFs and explain how the application of this scheme to $\mathbb{Z}_2$-odd
insulators is different than for ordinary insulators. The localization
properties of the constructed WFs are described in Sec.~\ref{Sec.5}.
The electric polarization properties and locations of the Wannier charge
centers are considered in Sec.~\ref{Sec.6}. Finally, we make concluding
remarks in Sec.~\ref{Sec.7}.

\section{\label{Sec.2} $\mathbb{Z}_2$ invariant}

Here we briefly review some of the equivalent ways of
determining the $\mathbb{Z}_2$ invariant in 2D insulators.

\begin{figure}
\begin{center}
\includegraphics[width=2.8in, bb= 7 13 276 276]{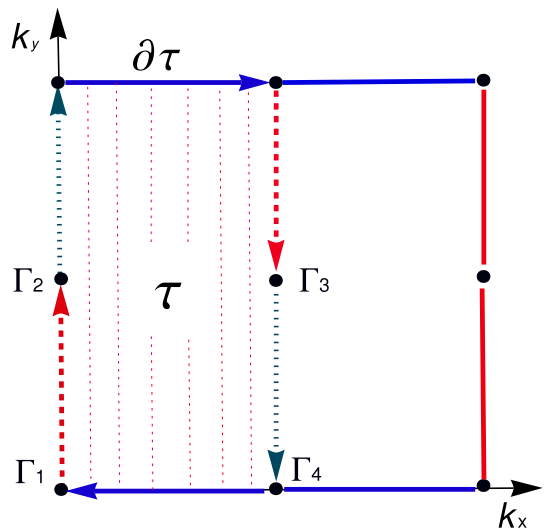}
\end{center}
\caption{(Color online) Sketch of the Brillouin zone. The Berry curvature of
Eq.~(\ref{Z2def-3}) is calculated in the interior of the half zone
$\tau$ (dashed region), while the Berry connection is evaluated
along its boundary $\partial \tau$ (arrows indicate direction
of integration).  Time-reversal--invariant points $\Gamma_i$
are shown.}
\label{bz}
\end{figure}

In the work of Ref.~\onlinecite{Kane-PRL05-a}
the definition of the $\mathbb{Z}_2$
invariant was given in terms of a function $P({\bf k})$ defined as
\begin{equation}
P({\bf k})=\mathrm{Pf}[\langle u_i({\bf k})|\hat{\theta} |u_j({\bf k})\rangle],
\label{Pfaffian}
\end{equation}
i.e.,  the Pfaffian of a certain $\bf k$-dependent antisymmetric
$N\times N$ matrix, where $N$ is the number of occupied bands.
Here $|u_j({\bf k})\rangle=e^{-i\k\cdot\r}|\psi_j({\bf k})\rangle$
is the periodic part of the Bloch
function of the $j$'th occupied band
and $\hat{\theta}=is^y\hat{C}$
is the time-reversal operator ($\hat{C}$ is complex conjugation and
$s^y$ is the second Pauli matrix).
If the zeros of $P({\bf k})$ are discrete, then
the $\mathbb{Z}_2$ invariant is odd if
the number of zeros of the Pfaffian within one half of the
Brillouin zone (BZ) (see Fig.~\ref{bz})
is odd, and even otherwise.
If the zeros of the Pfaffian occur along lines in the
BZ, then the $\mathbb{Z}_2$ invariant depends similarly on
whether half the number of sign changes of
$P({\bf k})$ along the boundary of the half BZ is odd or
even.  Using $\Delta=0$ and 1 to represent evenness and oddness
respectively, the $\mathbb{Z}_2$ invariant can equivalently be
determined as \cite{Kane-PRL05-b}
\begin{equation}
\Delta=\frac{1}{2i\pi}\oint_{\partial \tau} d{\bf k}\cdot \nabla_{\bf k}
\log[P({\bf k}+ i \delta)] \mod{2},
\label{Z2def-1}
\end{equation}
where the loop integral runs along the boundary $\partial \tau$ of the
half BZ, and the $\delta$ term is included for convergence.

Another approach to the problem of defining $\Delta$ results from
considerations of ``time-reversal polarization.''\cite{Fu-PRB06}   Here
a spin-pumping cycle is considered and it is shown that the
$\mathbb{Z}_2$ index
is given by the difference between the time-reversal polarizations at the
beginning and the midpoint of the cycle. This approach leads to the formula
\begin{equation}
(-1)^{\Delta} =\prod_{i=1}^4
\frac{\sqrt{\det[w({\bf \Gamma}_i)]}}{\mathrm{Pf}[w({\bf \Gamma}_i)]},
\label{Z2def-2}
\end{equation}
where $w_{mn}({\bf k}) = \langle u_m(-{\bf k})|\hat{\theta}| u_n({\bf k})
\rangle$ and ${\bf \Gamma}_i$ are the four time-reversal invariant
points of the BZ (i.e., those for which $-{\bf \Gamma}_i=
{\bf \Gamma}_i+{\bf G}$ with $\bf G$ a reciprocal vector). Note that
the matrix $w_{mn}$ is not the same as that in Eq.~(\ref{Pfaffian}).

The definition in Eq.~(\ref{Z2def-2}) appears to require a knowledge
of the occupied wavefunctions at only four points in the BZ, unlike
Eq.~(\ref{Z2def-1}), for which the wavefunctions must be known at
all points along the boundary of the half BZ. However, Eq.~(\ref{Z2def-2})
is usually not suitable for numerical implementation in practice,
since the sign of the Pfaffian at any one of the four points can be
flipped by a relabeling of the Kramers-degenerate states at that point.
To be more explicit, there is a ``gauge freedom'' in the choice of
states $|u_m({\bf k})\rangle$, corresponding to a $\bf k$-dependent
$N\times N$ unitary rotation among the occupied states.
Eq.~(\ref{Z2def-2}) is only meaningful when a globally smooth gauge
choice enforces a relation between the labels at the four special
$\bf k$-points.\cite{Fu-PRB06}
This problem may be avoided in the presence of some additional symmetry
that can be used to establish the labels of the bands at these points.
For example, in Ref.~\onlinecite{Fu-PRB07} it is shown how
the presence of inversion symmetry allows for a simplified
calculation of $\Delta$ from Eq.~(\ref{Z2def-2}).

In the absence of inversion symmetry, one can use yet another
definition of the $\mathbb{Z}_2$ index taking the form \cite{Fu-PRB06}
\begin{equation}
\Delta=\frac{1}{2\pi} \left[ \oint_{\partial \tau} {\cal A}  d\ell  -
\int_\tau {\cal F} d\tau \right] \mod 2,
\label{Z2def-3}
\end{equation}
where ${\cal A}=i\sum_{n=1}^{\cal N} \langle u_n |\nabla_{\bf k}| u_n \rangle$
is the Berry connection of $\cal N$ occupied states and ${\cal F}=
\nabla_{\bf k} \times {\cal A}$ is the corresponding Berry
curvature.\cite{Berry-PRSL84}
Of course, if $\cal A$ and $\cal F$ are both constructed from a
common gauge that is smooth over $\tau$, the result would
vanish by Stokes' theorem.  Thus, Eq.~(\ref{Z2def-3}) is only made
meaningful by the additional specification \cite{Fu-PRB06}
that the boundary integral of $\cal A$ must be calculated using
a gauge that respects time-reversal symmetry, i.e.,
\begin{eqnarray}
|u_{2n-1}(-{\bf k})\rangle=\hat{\theta} |u_{2n}({\bf k})\rangle, \nonumber\\
|u_{2n}(-{\bf k})\rangle=-\hat{\theta} |u_{2n-1}({\bf k})\rangle.
\label{constraint2}
\end{eqnarray}
For the case of the nontrivial $\mathbb{Z}_2$ state, it turns out
to be impossible to choose a gauge that satisfies both smoothness
over $\tau$ and the constraint (\ref{constraint2}) over $\partial\tau$.
In other words, $\Delta$=1 signals the existence of the topological
obstruction.

To see how this works more explicitly, the contributions to the
integral of $\cal A$ over $\partial\tau$ are illustrated in Fig.~\ref{bz}.
We choose a gauge that is periodic, $|u_j(\k)\rangle=|u_j(\k+\G)\rangle$,
in addition to satisfying Eq.~(\ref{constraint2}).
The contributions of the top and bottom segments (solid blue arrows
in Fig.~\ref{bz}) then cancel because they are connected by
a reciprocal lattice vector $\G$.
Thus, the gauge needs to be fixed only along the left
and right boundaries (composed of red dashed and gray dotted arrows
in Fig.~\ref{bz}), which are separated by a half
reciprocal lattice vector.
At each of the special points $\Gamma_i$,
one state from each Kramers-degenerate pair is arbitrarily identified
as $|u_{2n-1}(\Gamma_i)\rangle$, and the other is constructed via
\begin{equation}
|u_{2n}(\Gamma_i)\rangle=-\hat{\theta}|u_{2n-1}(\Gamma_i)\rangle.
\label{constraint_2}
\end{equation}
Then we can make an arbitrary gauge choice along the remaining
portions of the gray dotted arrows in Fig.~\ref{bz} -- e.g., accepting
the output of some numerical diagonalization procedure.  Finally,
the gauge should be transferred to the dashed-arrow segments using
Eq.~(\ref{constraint2}),
where ${\bf k}$ and $-{\bf k}$ belong to the dotted and dashed segments
respectively.

Eq.~(\ref{Z2def-3}) can now be evaluated using a uniform discretized mesh
$\mathbb{K}$ covering the region $\tau$, with the time-reversal
constraint applied to the boundary $\partial\tau$ as described above.
To do so, define the link matrices
$M_{\mu,nm}({\bf k})=\langle u_n({\bf k})|u_m({\bf k}+{\bf s}_\mu)\rangle$
and the unimodular link variables
$L_\mu({\bf k})=\det M_\mu/|\det M_\mu|$,
where ${\bf k}\in \mathbb{K}$ and ${\bf s}_1$ (${\bf s}_2$) is
the step of the mesh in the direction of the reciprocal lattice
vector ${\bf G}_1$ (${\bf G}_2$). By defining
$A_1({\bf k})=\log L_1({\bf k})$ and
\begin{equation}
F({\bf k})=\log[L_1({\bf k})
L_2({\bf k}+{\bf s}_1) L_1^{-1}({\bf k}+{\bf s}_2) L_2^{-1}({\bf k})],
\label{FFF}
\end{equation}
one can write the lattice definition of the $\mathbb{Z}_2$ invariant as
\begin{equation}
\Delta_L=\frac{1}{2i \pi}\left[ \sum_{{\bf k}\in \partial \tau}
A_1({\bf k}) - \sum_{{\bf k}\in \tau} F ({\bf k})\right] \mod 2.
\label{Lattice-def}
\end{equation}
For a sufficiently fine mesh there will be no ambiguity in the branch
choice for the complex log in Eq.~(\ref{FFF}), since the argument of
the log must approach unity as the mesh becomes dense.  Moreover,
a change in the branch choice determining one of the boundary links
$A_s({\bf k})$ has no effect (mod 2) on Eq.~(\ref{FFF}),
since each $A_s({\bf k})$ appears twice as a result of the gauge-fixing
on the boundary.
Thus, once the mesh is fine enough so that the branch choices in
Eq.~(\ref{FFF}) are all unambiguous, Eq.~(\ref{Lattice-def}) gives
$\Delta$ exactly.\cite{Fukui-JPSJ07}

\section{\label{Sec.3} The Kane-Mele model}

In their remarkable paper introducing a $\mathbb{Z}_2$ topological
classification to distinguish a QSH ($\mathbb{Z}_2$-odd) insulator
from an ordinary ($\mathbb{Z}_2$-even) insulator, 
Kane and Mele (KM) \cite{Kane-PRL05-b}  also
introduced a model tight-binding Hamiltonian that describes a 
2D $\mathbb{Z}_2$-odd 
insulator in some of its parameter space.  In this section we will
describe some of the properties of the model suggested therein.

The KM model is a tight-binding model on a honeycomb lattice with
one spinor orbital per site.  The primitive hexagonal lattice vectors
are ${\bf a}_{1,2}=a/2(\sqrt{3}\hat{\bf y}\pm \hat{\bf x})$
and sites $A$ and $B$ are located at
${\bf t}_A=a{\hat{y}}/\sqrt{3}$ and ${\bf t}_B=2a{\hat{y}}/\sqrt{3}$
respectively. The KM Hamiltonian is
\begin{eqnarray}
 H&=&t\sum_{<ij>} c_i^\dagger c_j
   + i \lambda_{\rm SO} \sum_{\ll ij \gg} \nu_{ij} c_i^\dagger s^z c_j 
\nonumber \\
  &+& i \lambda_{\rm R} \sum_{<ij>} c_i^\dagger ({\bf s} \times
    \hat{\bf d}_{ij})_z c_j+\lambda_v\sum_{i}\xi_i c_i^\dagger c_i, 
\label{second-quantKM}
\end{eqnarray}
where the spin indices have been suppressed on the raising and lowering
operators, and 
$t$ is the nearest-neighbor hopping amplitude.  In the second term,
$\lambda_{\rm SO}$ is the strength of the spin-orbit interaction acting
between second neighbors, with $\nu_{ij}=(2/\sqrt{3})[\hat{\bf
d}_1 \times \hat{\bf d}_2]= \pm 1 $ depending on the relative
orientation of the first-neighbor bond vectors $\hat{\bf d}_1$
and $\hat{\bf d}_2$ encountered by an electron hopping from site
$j$ to site $i$, and $s^z$ is the $z$ Pauli spin matrix.  Next,
$\lambda_{\rm R}$ describes the Rashba interaction\cite{Bychkov-JETP84}
that couples differently oriented first-neighbor spins, with ${\bf
s}$ being the vector of Pauli matrices.  Finally, $\lambda_v$
is the strength of the staggered on-site potential, for which
$\xi_i$ is $+1$ and $-1$ on A and B sites respectively.  Note that
the symmetry of the problem is lowered significantly compared to
an ideal honeycomb lattice, since the on-site staggered potential
makes the A and B sites inequivalent, 
while the Rashba term breaks $s^z$ conservation.

\begin{figure}
\begin{center}
\includegraphics[width=2.8in]{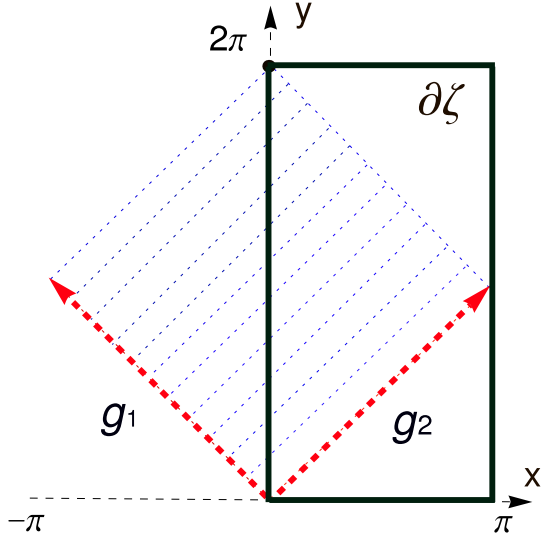}
\end{center}
\caption{(Color online) Brillouin zone sketched using coordinates
$x=k_xa/2$ and $y=\sqrt{3}k_ya/2$.  Primitive
reciprocal lattice vectors
${\bf G}_1=(2\pi/a)(1,1/\sqrt{3})$ and
${\bf G}_2=(2\pi/a)(-1,1/\sqrt{3})$
correspond to ${\bf g}_1=(\pi,\pi)$ and ${\bf g}_2=(-\pi,\pi)$
respectively.  The black rectangle marks the boundary
$\partial \zeta$ of the zone used for polarization calculations
in Sec.~\ref{Sec.6}.}
\label{ebz}
\end{figure}

To proceed, we choose the tight-binding basis wavefunctions to be
\begin{equation}
\chi_{j\sigma{\bf k}}({\bf r})=(1/\sqrt{N}) \sum_{\bf R}
e^{i {\bf k} \cdot {\bf R}} \phi_\sigma({\bf r}-{\bf R}-{\bf t}_j),
\label{TBwf}
\end{equation}
where $\sigma$ is a spin index,
$j=\{A,B\}$ denotes the atom type, ${\bf t}_j$ is a vector
that specifies the position of the atom in the unit cell,\footnote{
  An alternative tight-binding convention, defined
  by replacing $e^{i\k\cdot\R}$ by $e^{i\k\cdot(\R+{\bf t}_j)}$
  in Eq.~(\ref{TBwf}), is possible but is not adopted here.}
and ${\bf R}$ is a lattice vector built from the primitive lattice vectors
${\bf a}_1$ and ${\bf a}_2$. This allows the Hamiltonian to be written as
a 4$\times$4 matrix $H_{j\sigma,j'\sigma'}({\bf k})=\langle
\chi_{j\sigma{\bf k}} \vert H \vert \chi_{j'\sigma'{\bf k}} \rangle$,
which can be cast in terms of
five Dirac matrices $\Gamma^{\alpha}$ and their ten commutators
$\Gamma^{\alpha \beta}=[\Gamma^ {\alpha},\Gamma^{\beta}]/(2i)$
as
\begin{equation}
H({\bf k})=\sum_{\alpha=1}^5 d_{\alpha} ({\bf k}) \,\Gamma^{\alpha} +
\sum_{\alpha<\beta=1}^5 d_{\alpha \beta} ({\bf k}) \,\Gamma^{\alpha \beta}
\label{Dirac-Hamiltonian}
\end{equation}
where the Dirac matrices are chosen to be $\Gamma^{1,2,3,4,5} =
(I\otimes \sigma^x, I\otimes \sigma^z, s^x \otimes \sigma^y,
s^y \otimes \sigma^y, s^z \otimes \sigma^y)$ with the Pauli
matrices $\sigma^k$ and $s^k$ acting in sublattice and spin
space respectively.
The dependence of the $d_\alpha$ and $d_{\alpha\beta}$
coefficients on wavevector is detailed in Table~\ref{Tab:coefficients}
using the notation $x=k_xa/2$ and $y=\sqrt{3}k_ya/2$, with the
relationship of these variables to the BZ being sketched
in Fig.~\ref{ebz}.

\begin{table}
\begin{center}
\begin{tabular}{c c c c}
\hline
\hline
$d_1$ & $t(1+2\cos{x}\cos{y})$               &
$d_{12}$ & $ -2t\cos{x}\sin{y}$                      \\
$d_2$ & $\lambda_v$                          &
$d_{15}$ & $2\lambda_{\rm SO}(\sin{2x}-2\sin{x}\cos{y})$ \\
$d_3$ & $\lambda_{\rm R}(1-\cos{x}\cos{y})$        &
$d_{23}$ & $-\lambda_{\rm R}\cos{x}\sin{y}$                 \\
$d_4$ & $-\sqrt{3}\lambda_{\rm R} \sin{x}\sin{y}$ &
$d_{24}$ & $\sqrt{3}\lambda_{\rm R} \sin{x}\cos{y}$         \\
\hline
\hline
\end{tabular}
\end{center}
\caption{(Color online) Nonzero coefficients appearing in
Eq.~(\ref{Dirac-Hamiltonian}), using the notation
$x=k_xa/2$ and $y=\sqrt{3}k_ya/2$ (see also Fig.~\ref{ebz}).}
\label{Tab:coefficients}
\end{table}

Since, $\hat{\theta} \Gamma^{\alpha} \hat{\theta}^{-1}=\Gamma^{\alpha}$
and $\hat{\theta} \Gamma^{\alpha \beta} \hat{\theta}^{-1}=-\Gamma^{\alpha
\beta}$, while $d_\alpha({\bf k})=d_\alpha(-{\bf k})$ and $d_{\alpha \beta}
({\bf k}) = - d_{\alpha \beta}(-{\bf k})$, the Hamiltonian
(\ref{second-quantKM}) is time-reversal invariant, i.e.,
$\hat{\theta} H({\bf k}) \hat{\theta}^{-1} = H(-{\bf k})$.  However,
it lacks particle-hole symmetry in the sense of
Refs.~(\onlinecite{Schnyder-PRB08,Schnyder-AIP09,Kitaev-AIP09}), because
of the action of the on-site and spin-orbit coupling terms.
In the general classification of topological insulators and
superconductors,\cite{Schnyder-PRB08, Schnyder-AIP09,Kitaev-AIP09}
therefore, the Kane-Mele model falls into the AII symplectic symmetry
class, which in two dimensions has a $\mathbb{Z}_2$ classification.
This means that by varying parameters of the Hamiltonian of
Eq.~(\ref{second-quantKM}) one can switch between $\mathbb{Z}_2$-odd
and $\mathbb{Z}_2$-even phases, with the system experiencing a gap
closure and becoming metallic at the transition from one phase to the other.

\begin{figure}
\begin{center}
\includegraphics[width=2.4in, bb=2 2 245 245]{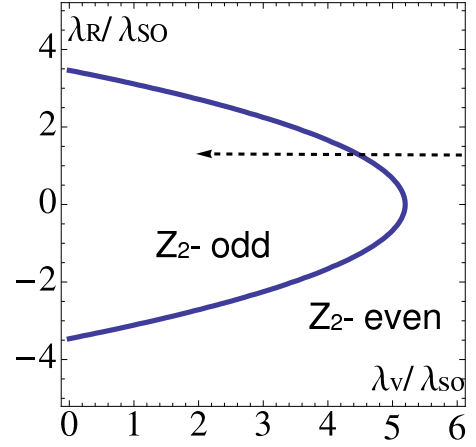}
\end{center}
\caption{Phase diagram of the Kane-Mele model for
$\lambda_v/\lambda_{\rm SO}$ $>$ 0. Arrow illustrates a path crossing
the phase boundary by varying
$\lambda_v$ while keeping other parameters fixed.}
\label{PhaseD}
\end{figure}

For the present purposes we assume $\lambda_{\rm SO}>0$ without
loss of generality. We also fix $\lambda_v>0$. For this case, the
transition between $\mathbb{Z}_2$-odd and $\mathbb{Z}_2$-even phases
is accompanied by a gap closure at the $K$ and $K'$ 
points (the zone-boundary points of three-fold symmetry) in the BZ.
The energy is independent of $t$ at these points, and $\lambda_{\rm SO}$
can be used as the energy
scale. The energy gap is then given by $|6\sqrt{3}-\lambda_v/\lambda_{\rm SO}-
\sqrt{(\lambda_v/\lambda_{\rm SO})^2+9(\lambda_{\rm R}/\lambda_{\rm SO})^2}|$, leading to
the phase diagram shown in Fig.~\ref{PhaseD}.

Note that when $\lambda_{\rm R}=0$ the model reduces to two independent copies
of the Haldane model\cite{Haldane-PRL88}
the $\mathbb{Z}_2$ invariant is odd
when the Chern numbers are odd, and even otherwise.\cite{Sheng-PRL06}

In what follows we use $t$ as the energy scale and fix the values
of the other parameters to be $\lambda_{\rm SO}/t=0.6$ and $\lambda_{\rm R}/t=0.5$.
Varying the third parameter $\lambda_v/t$ allows us to switch from
the $\mathbb{Z}_2$-even to the $\mathbb{Z}_2$-odd phase. The phase
transition occurs at $|\lambda_v|/t\simeq 2.93$, with the system in
the $\mathbb{Z}_2$-odd phase for $-2.93<\lambda_v/t<2.93$.
As discussed above, the energy gap closes at the phase transition, and
remains open in both the $\mathbb{Z}_2$-odd and $\mathbb{Z}_2$-even
phases.

\section{\label{Sec.4} Gauge freedom and Wannier functions}

\subsection{\label{gen-consid} General considerations}

We now consider the problem of constructing Wannier functions (WF)
for the Kane-Mele model.  We emphasize that we mean by this a set
of localized functions spanning the same space as the occupied
Bloch bands.  Several recent papers have discussed the construction
of WFs for an enlarged subspace including also some unoccupied bands
for 3D topological insulators such as Bi$_2$Se$_3$,\cite{Zhang-NatPhys09,
Zhang-NJP10}
in which case there is typically no topological obstruction,
but this is not the context of the present work.

We start with the general definition of the WF in cell
${\bf R}$ and with band index $n$ in 2D,
\begin{equation}
\langle {\bf r}|{\bf R}n\rangle \equiv W_n({\bf r}-{\bf R}) =
\frac{A}{(2\pi)^2} \int_{BZ} d{\bf k}\, e^{-i {\bf k}\cdot{\bf R}}
\psi_{n{\bf k}}({\bf r}),
\label{WFdef}
\end{equation}
where $A$ is the unit cell area and Bloch wavefunctions $\psi_{n{\bf k}}$ are
assumed to be normalized within the unit cell. This definition is not
unique; not only is there the usual ${\cal U}(1)$ gauge freedom associated
with a $\bf k$-dependent phase twist of each band $n$, there is more
generally a ${\cal U}({\cal N})$
gauge freedom
\begin{equation}
\vert\psi_{n{\bf k}}\rangle \; \longrightarrow \; \sum_m U_{mn}({\bf k}) \,
\vert\psi_{m{\bf k}}\rangle
\label{Ugauge}
\end{equation}
coming from the fact that the $\cal N$ occupied Bloch bands can be mixed
with each other by a $\bf k$-dependent ${\cal U}({\cal N})$ transformation.
In fact, it is generally necessary to pre-mix the Bloch states
using this ${\cal U}(N)$ gauge freedom in order that the resulting
Bloch-like states (and their phases) will be smooth functions of
$\bf k$.  However, having done so, there is still a large gauge
freedom associated with the application of a subsequent ${\cal U}({\cal N})$
gauge rotation that is smooth in $\bf k$.

This ambiguity in the gauge choice can be removed by applying some
criterion to the selection of the WFs. Since electrons are expected to be
localized in insulators,\cite{Resta-PRL99} a sensible criterion is that
of Ref.~\onlinecite{Marzari-PRB97}, which specifies maximal localization
of the WFs in real space.
In this approach, which we adopt
here, one chooses some localized trial functions in order to provide
a starting guess about where the electrons are localized in the unit
cell, and obtains a fairly well-localized set of WFs by a projection
procedure to be described shortly.  If desired, one can follow this
with an iterative procedure to make the resulting WFs optimally
localized.\cite{Marzari-PRB97}

Consider an insulator with $\cal N$ occupied bands.  We start with
a set of $\cal N$ trial states $|\tau_{i}\rangle$ located in the
home unit cell, and at each $\bf k$ we project them onto the
occupied subspace at $\bf k$ to get a set of Bloch-like states
\begin{equation}
|\Upsilon_{i{\bf k}}\rangle=\hat{P}_{\bf k}\,|\tau_i\rangle
= \sum_{n=1}^{\cal N}
|\psi_{n{\bf k}} \rangle \langle \psi_{n{\bf k}}| \tau_i \rangle.
\label{Upsilon}
\end{equation}
Since this set of states will not generally be orthonormal,
we make use of a L\"owdin orthonormalization procedure which
consists of constructing the overlap matrix
\begin{equation}
S_{mn}({\bf k})=\langle \Upsilon_{m{\bf k}} |\Upsilon_{n{\bf k}} \rangle
\label{S-matrix}
\end{equation}
and obtaining the orthonormal set of Bloch-like orbitals
\def\bls{\tilde{\psi}_{n{\bf k}}}
\begin{equation}
 \vert \bls \rangle = \sum_m \left[S({\bf k})^{-1/2} \right]_{mn}
|\Upsilon_{m{\bf k}} \rangle.
\label{realLWF}
\end{equation}
Note that the $\bls$ are not eigenstates of the Hamiltonian,
but they span the same space, and have the same form, as the
usual Bloch eigenstates.
For an insulator whose gap is not
too small, and for a set of trial functions embodying a reasonable
assumption about character of the localized electrons,
the $\bls$ will be smooth functions of $\bf k$.  In that
case, by the usual properties of Fourier transforms, the
WFs constructed in analogy with Eq.~(\ref{WFdef}),
\begin{equation}
|{\bf R}n\rangle =
\frac{A}{(2\pi)^2} \int_{BZ} d{\bf k}\, e^{-i {\bf k}\cdot{\bf R}}
\, \vert \bls \rangle ,
\label{WFdefg}
\end{equation}
should be well localized.

Such a construction will break down if the
determinant of $S({\bf k})$ vanishes at any $\bf k$.
This is guaranteed to occur in a Chern insulator, where time-reversal
symmetry is broken and the Chern index of the occupied manifold is
non-zero; in this case, construction of exponentially localized WFs
becomes impossible.\cite{Brouder-PRL07,Thonhauser-PRB06,Thouless-JPC84}
For a $\mathbb{Z}_2$ insulator, however, the presence of time-reversal
symmetry guarantees a zero Chern index, so that exponentially
localized WFs must exist.\cite{Brouder-PRL07}
In this case, we should be able to find a set of
trial functions such that ${\rm det}\,S({\bf k})\ne0$
throughout the BZ.

\subsection{\label{Sec4Sub1} $\mathbb{Z}_2$-even phase}

\begin{figure}
\begin{center}
\includegraphics[width=3.2in, bb=34 52 737 588]{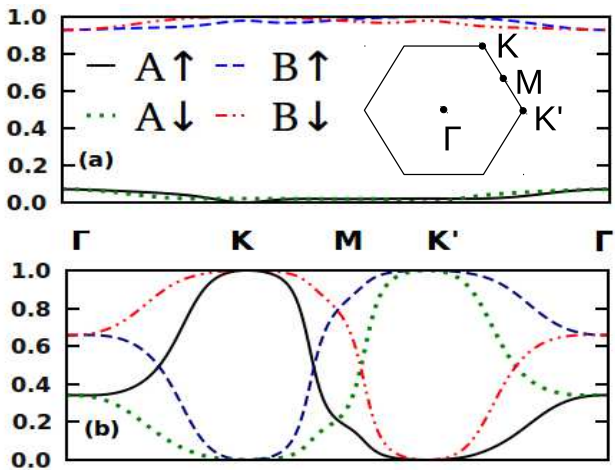}
\end{center}
\caption{(Color online) Sum of the weights of the projections
into the two occupied bands of the basis states
\mbox{$|A;\uparrow_z\rangle$},
\mbox{$|B;\uparrow_z\rangle$},
\mbox{$|A;\downarrow_z\rangle$}, and
\mbox{$|B;\downarrow_z\rangle$}
plotted along the diagonal of the BZ for (a) $\lambda_v/t=5$ ($\mathbb{Z}_2$-even
phase) and (b) $\lambda_v/t=1$ ($\mathbb{Z}_2$-odd phase). 
Inset in (a): BZ of a honeycomb lattice.}
\label{wcomb}
\end{figure}

\begin{figure}
\begin{center}
\includegraphics[width=3.2in, bb= 3 15 350 278]{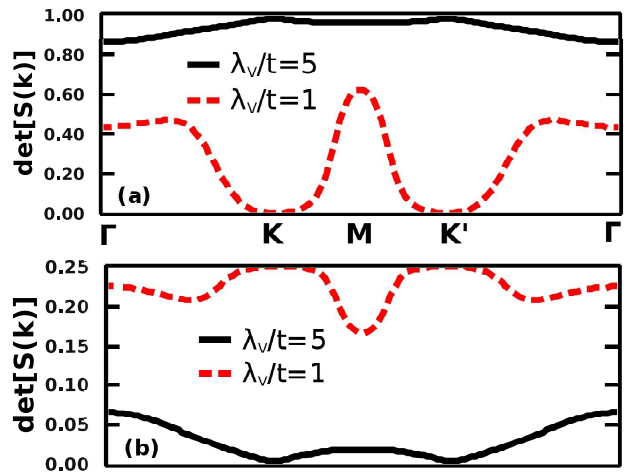}
\end{center}
\caption{(Color online) Plot of $\det[S({\bf k})]$ along the diagonal
of the BZ
for $\lambda_v/t=5$ ($\mathbb{Z}_2$-even phase) and $\lambda_v/t=1$
($\mathbb{Z}_2$-odd phase). (a) Trial functions are
$|B;\uparrow_z\rangle$ and
$|B;\downarrow_z\rangle$.
(b) Trial functions are
$|A;\uparrow_x\rangle$ and
$|B;\downarrow_x\rangle$.}
\label{dets}
\end{figure}

Let us first apply the method described above to the case of the
$\mathbb{Z}_2$-even
phase of the Kane-Mele model. This phase is topologically equivalent to the
ordinary insulator, so we anticipate a picture in which the two
electrons per cell are opposite-spin ones approximately localized on
the lower-energy ($B$) site.
One way to see this is to look at the weights of the
basis states in the occupied subspace.
Figure~\ref{wcomb}(a) shows the
distribution of these weights along a high-symmetry line in the BZ for
the Kane-Mele model in its $\mathbb{Z}_2$-even phase.  From the figure 
it is obvious that the two basis states on the $B$ site
dominate in the occupied subspace over the whole BZ.  It is
then natural to choose the two trial functions to be opposite-spin
spatial $\delta$-functions localized on the $B$ site in the home
unit cell.  We choose these to be spin-aligned along $z$, i.e.,
\begin{equation}
|\tau_i\rangle = |B;\sigma^z_i\rangle =
   \delta({\bf r}-{\bf t}_B)|\sigma^z_i\rangle
\label{trial-1}
\end{equation}
where $|\sigma^z_1\rangle=\vert\!\!\uparrow_z\rangle$
and $|\sigma^z_2\rangle=|\!\!\downarrow_z\rangle$.
Transforming to ${\bf k}$-space we get
\begin{equation}
|\tau_{i{\bf k}}\rangle = \frac{|\sigma^z_i\rangle}{\sqrt{N}}\sum_{\bf R}
e^{i {\bf k} \cdot {\bf R}}\delta({\bf r}-{\bf R}- {\bf t}_B) .
\label{trial-1k}
\end{equation}
The two occupied Bloch bands may be written as
\begin{equation}
|\psi_{n{\bf k}}\rangle=\sum_{\ell} C_{\ell n{\bf k}}|\chi_{\ell{\bf k}}
\rangle
\label{Bloch-bands}
\end{equation}
where $\ell$ is a combined index for sublattice and spin, $\ell=\{1,2,3,4\}
\equiv \{A \uparrow, B \uparrow, A\downarrow, B \downarrow\}$, and
$\chi_{\ell {\bf k}}=\chi_{j \sigma {\bf k}}$ are the tight-binding basis
functions of Eq.~(\ref{TBwf}).
With Eq.~(\ref{trial-1k}) the projected functions become
\begin{equation}
|\Upsilon_{1{\bf k}}\rangle 
   =C^*_{21{\bf k}} |\psi_{1{\bf k}} \rangle + C^*_{22{\bf k}}
|\psi_{2{\bf k}} \rangle,
\label{Ups-1}
\end{equation}
\begin{equation}
|\Upsilon_{2{\bf k}}\rangle 
  =C^*_{41{\bf k}} |\psi_{1{\bf k}} \rangle + C^*_{42{\bf k}}
|\psi_{2{\bf k}} \rangle.
\label{Ups-2}
\end{equation}
The overlap matrix $S$ is constructed from these functions, and for
the determinant one finds
\begin{eqnarray}
\det{[S({\bf k})]}&=&(|C_{21{\bf k}}|^2 + |C_{22{\bf k}}|^2) (|C_{41{\bf k}}|^2 +
|C_{42{\bf k}}|^2)  \nonumber \\
&-& |C_{21{\bf k}} C_{41{\bf k}}^*+ C_{22{\bf k}}
C_{42{\bf k}}^*|^2   .
\label{det-1}
\end{eqnarray}
Recall that for the L\"owdin orthonormalization procedure to succeed,
this determinant must remain non-zero everywhere in the BZ.  This
is indeed the case for the $\mathbb{Z}_2$-even phase, as illustrated in
Fig.~\ref{dets}(a),
where the solid black curve shows the dependence
of the determinant on ${\bf k}$ along the high-symmetry line in the BZ.

In contrast, the dashed red curve in Fig.~\ref{dets}(a)
shows the
behavior of $\det{[S({\bf k})]}$ in the $\mathbb{Z}_2$-odd regime.
The determinant can be seen to vanish at the $K$ and $K'$ points in
the BZ.  Clearly, this choice of trial functions is not appropriate
for building the Wannier representation in the $\mathbb{Z}_2$-odd phase.
Indeed, as we shall see in the next subsection,
{\it any} choice of trial functions that come in Kramers pairs
is guaranteed to fail in the $\mathbb{Z}_2$-odd case.
There we shall also investigate alternative choices of trial
functions that allow for a successful construction of WFs.

\subsection{\label{Sec4Sub2} $\mathbb{Z}_2$-odd phase}

To gain some insight into the appropriate choice of trial functions
in the $\mathbb{Z}_2$-odd regime, consider the weights of the
basis functions in the occupied space shown for this case in
Fig.~\ref{wcomb}(b).
Unlike the normal insulator, the $\mathbb{Z}_2$-odd phase does not
favor any particular basis states.  Instead, different basis states
dominate in different portions of the BZ.
For example, at points $K$ and $K'$ the occupied space is
represented by only two of the four basis states; at each of
these points the two participating basis states have opposite spin
and sublattice indices, and none appear in common at both points.
(The states at $K$ are, of course, Kramers pairs of those at $K'$.)
It follows that if any of the trial states is simply set equal to
one of the four basis states, then at least one of the
$|\Upsilon \rangle$ would vanish either at $K$ or $K'$, and
the determinant would vanish there too.  This explains the
failure of the naive Wannier construction procedure for the
$\mathbb{Z}_2$-odd phase; with the naive choice of trial
functions as in Eq.~(\ref{trial-1}), the determinant vanishes at
both $K$ and $K'$, as shown by the red dashed curve in
Fig.~\ref{dets}(a).\footnote{Fig.~\ref{wcomb}(b) also shows
that the character of the occupied states changes in the BZ, 
which serves as an illustration of the band inversion
associated with topological insulators.}

In fact, this failure can be understood from a general point of view.
If the two trial functions form a Kramers pair, then the projection
procedure of Eqs.~(\ref{Upsilon}-\ref{realLWF}) will result in
Bloch-like functions obeying
\begin{eqnarray}
|\tilde{\psi}_1(-{\bf k})\rangle=\theta |\tilde{\psi}_2({\bf k})\rangle,
\nonumber\\
|\tilde{\psi}_2(-{\bf k})\rangle=-\theta |\tilde{\psi}_1({\bf k})\rangle.
\label{constraint}
\end{eqnarray}
The WFs obtained from Eq.~(\ref{WFdefg}) will then
also form a Kramers pair.
But Eq.~(\ref{constraint}) is nothing other than
the constraint of Eq.~(\ref{constraint2}) defining a gauge
that respects time-reversal symmetry, and it has been shown
\cite{Fu-PRB06,Roy-PRB09-a,Loring-arx10} that an odd value of the $\mathbb{Z}_2$
invariant presents an obstruction against constructing such
a gauge.  In other words,
in the $\mathbb{Z}_2$-odd phase a smooth gauge cannot be fixed
by choosing trial functions that are time-reversal pairs of
each other, and a choice of WFs as time-reversal
pairs is not possible. Hence, in order to construct the Wannier
representation in the $\mathbb{Z}_2$-odd regime, one should
choose trial functions that do not transform into one another
under time reversal.

Following these arguments, we choose the two trial functions to be
localized on {\it different} sites in the home unit cell.  Moreover,
in order that they will have components on states with spins both
up and down along $z$, we choose the spins of the trial states
so that one is along $+x$ and the other along $-x$.\footnote{
To see that the $z$ components have to be mixed, consider two trial
functions that are localized on different sites $A$ and $B$ with opposite
direction of spin in the $z$ direction. In this case the projected
functions $\Upsilon_{i\k}$ become zero either at $K$ or $K'$, as follows
from the Fig.~\ref{wcomb}(b), and $\det[S]$ becomes zero.}
In $\bf k$-space this becomes
\begin{equation}
|\tau_{i\bf k}\rangle = \frac{|\sigma^x_i\rangle}{\sqrt{N}}
\sum_{\bf R}e^{i {\bf k} \cdot {\bf R}}\delta({\bf r}-{\bf R}-{\bf t}_i)
\label{trial-2}
\end{equation}
where ${\bf t}_1={\bf t}_A$ and ${\bf t}_2={\bf t}_B$, leading to
\begin{equation}
|\Upsilon_{1{\bf k}} \rangle =\left[ (C_{11{\bf k}}^*+C_{31{\bf k}}^*)|\psi_1\rangle +
(C_{12{\bf k}}^*+C_{32{\bf k}}^*)|\psi_2\rangle \right] /{\sqrt{2}}
\label{upsilon1-2}
\end{equation}
and
\begin{equation}
|\Upsilon_{2{\bf k}} \rangle =\left[ (C_{21{\bf k}}^*-C_{41{\bf k}}^*)|\psi_1\rangle +
(C_{22{\bf k}}^*-C_{42{\bf k}}^*)|\psi_2\rangle \right] /{\sqrt{2}}.
\label{upsilon2-2}
\end{equation}
The determinant takes the form
\begin{widetext}
\begin{eqnarray}
\det[S]&=&(|C_{11{\bf k}}+C_{31{\bf k}}|^2+|C_{12{\bf k}}+C_{32{\bf k}}|^2)
(|C_{21{\bf k}}-C_{41{\bf k}}|^2+|C_{22{\bf k}}-C_{42{\bf k}}|^2)/4- \nonumber \\
&-&|(C_{11{\bf k}}+C_{31{\bf k}})(C_{21{\bf k}}^*-C_{41{\bf k}}^*)+
(C_{12{\bf k}}+C_{32{\bf k}}) (C_{22{\bf k}}^*-C_{42{\bf k}}^*)|^2/4.
\label{det-2}
\end{eqnarray}
\end{widetext}

The dependence $\det[S({\bf k})]$ is shown along the diagonal
of the Brillouin zone for this choice of trial functions in
Fig.~\ref{dets}(b). In the $\mathbb{Z}_2$-odd phase (dashed line)
the determinant remains non-zero everywhere in the BZ.\footnote{
Actually, the minimum value of $|\det\,S|$ occurs off the plotted
symmetry line and is $0.0873$.}  
Not surprisingly, the same trial
functions are very poorly suited to the normal-insulator phase,
as can be seen from solid line in the same panel.  In this case
$\det[S({\bf k})]$ almost vanishes at $K$ and $K'$ and remains
quite small throughout the rest of the BZ, so that one should
clearly revert to the time-reversed pair of trial functions
of Eq.~(\ref{trial-1}) and Fig.~\ref{dets}(a) in order to get
well-localized WFs.

We made an arbitrary choice above in selecting the two trial
functions to be up and down along $x$.  In fact, if we repeat
the entire procedure using trial functions that
are spin-up and spin-down along any unit vector $\hat{n}$ lying
in the $xy$-plane, we find that $\det[S({\bf k})]$ changes very
little, with only small changes in the size of the dip near the
$\Gamma$ point.  Thus, we find that the choice of trial functions
in Eq.~(\ref{trial-2})
is not unique. Instead, there is a large degree of arbitrariness
in the choice of WFs in the $\mathbb{Z}_2$-odd case.

To conclude, we have established that the choice of a time-reversal
pair of trial functions, Eq.~(\ref{trial-1}), that allows
for the construction of well-localized WFs in the ordinary-insulator
phase cannot be used in the $\mathbb{Z}_2$-odd phase. In order for the
usual projection method for constructing the Wannier representation
to work in this topologically nontrivial phase, the trial functions
should explicitly break time-reversal symmetry, i.e., they should
not come in time-reversal pairs.

\section{\label{Sec.5} Localization of Wannier Functions in the
$\mathbb{Z}_2$-odd Insulator}

Now that we know how to construct WFs for the $\mathbb{Z}_2$-odd
insulator, we discuss their localization
properties. As we have noted in the preceding section, the choice
of the trial functions, Eq.~(\ref{trial-2}), is not unique; there
are other gauge choices arising from different trial functions
that also produce well-defined sets of WFs.
Since different gauge choices
lead to different degrees of localization of the resulting WFs,
it is natural to fix the gauge by the condition of maximal possible
localization of the WFs.

The problem of constructing maximally-localized WFs was studied
by Marzari and Vanderbilt.\cite{Marzari-PRB97}  They
considered the total quadratic spread
\begin{equation}
\Omega=\sum_{n=1}^{\cal N} [ \langle {\bf 0}n|r^2|{\bf 0}n \rangle -
\langle {\bf 0}n | {\bf r} | {\bf 0}n \rangle^2 ]
\label{Spread-def}
\end{equation}
as a measure of the delocalization of WFs in real space, and
developed methods for iteratively reducing the spread via a
series of unitary transformations, Eq.~(\ref{Ugauge}),
applied prior to WF construction.
The spread functional was
decomposed into two parts, $\Omega=\Omega_I+
\tilde{\Omega}$, with
\begin{equation}
\Omega_I=\sum_{n=1}^{\cal N} \left[ \langle {\bf 0}n | r^2| {\bf 0}n \rangle -
\sum_{{\bf R}m} |\langle {\bf R}m|{\bf r}|{\bf 0}n \rangle |^2 \right]
\label{omegaI}
\end{equation}
being the gauge-invariant part and
\begin{equation}
\tilde{\Omega}=\sum_{n=1}^{{\cal N}} \sum_{{\bf R}m \neq {\bf 0}n}
| \langle {\bf R}m|{\bf r}|{\bf 0}n \rangle |^2
\label{omega-tilde}
\end{equation}
the gauge-dependent part of the spread.
Discretized ${\bf k}$-space
formulas for Eqs.~(\ref{omegaI}) and (\ref{omega-tilde}) were also
derived for the case that the BZ is represented by a uniform
$\bf k$ mesh.  The
resulting expression for the gauge-invariant spread is, for
example,
\begin{equation}
\Omega_I=\frac{1}{N}\sum_{{\bf k},{\bf b}} \omega_b \sum_{m,n=1}^{\cal N}
\left( \delta_{mn} - |M_{mn}^{({\bf k},{\bf k}+{\bf b})}|^2 \right),
\label{omegaI-k}
\end{equation}
where
\begin{equation}
M_{mn}^{({\bf k},{\bf k}+{\bf b})}=
\langle u_{n{\bf k}}|u_{m{\bf k}+{\bf b}}\rangle =
\sum_{\ell=1}^{4}C_{\ell n{\bf k}}^* C_{ \ell m{\bf k}+{\bf b}}
e^{-i {\bf b}\cdot {\bf t}_\ell}
\label{M-overlaps}
\end{equation}
are overlap matrices
and ${\bf b}$ are ``mesh vectors'' connecting each ${\bf k}$-point to
its nearest neighbors.  The latter are chosen, together with a set of
weights $\omega_b$, in such a way as to satisfy the condition
\begin{equation}
\sum_{\bf b}\omega_b b_i b_j = \delta_{ij}.
\label{weight-condition}
\end{equation}
A corresponding expression for $\tilde{\Omega}$, and
a description of steepest-descent methods capable of minimizing
$\Omega$, were also given in Ref.~\onlinecite{Marzari-PRB97}.
Note that, in order to avoid getting trapped in false local
minima, the iterative procedure is normally initialized using
the trial-function projection procedure described in Sec.~\ref{Sec.4}
above.

We now apply this method to the Kane-Mele model.
The lattice is hexagonal, and in
this case six ${\bf b}_j$ vectors are needed to satisfy the condition
(\ref{weight-condition}), namely
${\bf b}_1=-{\bf b}_4={\bf G}_1/q$,
${\bf b}_2=-{\bf b}_5=({\bf G}_1+{\bf G}_2)/q$, and
${\bf b}_3=-{\bf b}_6={\bf G}_2/q$.  All six have the
same length $b$ and weight $\omega_b=1/(3b^2)$.
We start with the WFs obtained with the projection
method using the trial functions of Eq.~(\ref{trial-2}),
appropriate for the $\mathbb{Z}_2$-odd phase.

\begin{figure}
\begin{center}
\includegraphics[width=3.4in, bb=14 24 786 356]{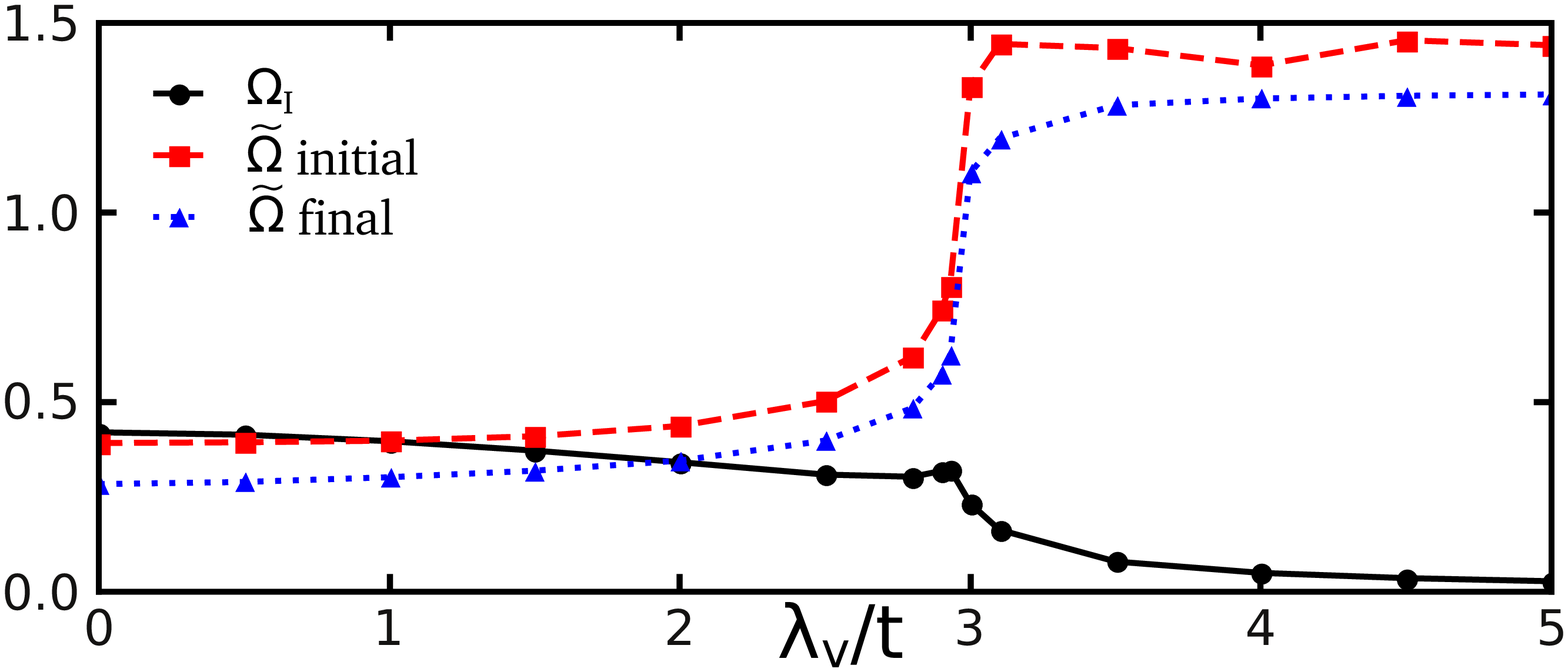}
\end{center}
\caption{(Color online)
Wannier spreads $\Omega_I$ and $\tilde{\Omega}$
for the Kane-Mele model on a 60$\times$60 ${\bf k}$-mesh,
initialized using the trial functions of Eq.~(\ref{trial-2}).
``Initial'' and ``final'' values are those computed before and
after the iterative minimization respectively.  The system is
in the $\mathbb{Z}_2$-odd phase for $\lambda_v/t\lesssim2.93$.}
\label{spread}
\end{figure}

The resulting
spreads, both before and after the iterative minimization,
are shown in Fig.~\ref{spread}.
($\Omega_I$, being gauge-invariant, is the same before and after.)
The left part of the figure shows the behavior in the
$\mathbb{Z}_2$-odd phase, where the trail functions are the
appropriate ones.
The results in this region were not strongly sensitive to the
$\bf k$-point mesh density.
The fact that $\tilde{\Omega}$ is similar in
magnitude to the unminimized $\Omega_I$, and that the localization
procedure reduces $\tilde{\Omega}$ by only $20-30\%$, provide
additional evidence that the choice of trial functions was a good
one.  The Wannier charge centers were almost unchanged by
the minimization procedure; the $x$-coordinates were zero, while
$\bar{r}_{1y}\simeq a/\sqrt{3}$ and $\bar{r}_{2y}\simeq 2a/\sqrt{3}$
(see Sec.~\ref{Sec.6} for details),
in good agreement with our initial assumption about the WFs being
localized on $A$ and $B$ sites.

The right part of Fig.~\ref{spread}, for $\lambda_v/t\gtrsim2.93$,
shows what happens when we attempt to use the same trial functions
in the normal phase.  $\Omega_I$ is of course unaffected by the
choice of trial functions, and the fact that it has a smaller
value in this region indicates, not surprisingly, that the
insulating state is simpler and more localized in the normal
state.  (For large $\lambda_v/t$ the WFs approach spatial
delta functions, explaining the fact that $\Omega_I$ asymptotes
to zero in that limit.)  Not surprisingly, however, using the
trial functions appropriate to the $\mathbb{Z}_2$-odd phase in the
$\mathbb{Z}_2$-even regime results in very poor localization of
the WFs as measured by $\tilde{\Omega}$.  Our data also suggests that in
the $\mathbb{Z}_2$-odd phase MLWFs are less localized than MLWS in
the $\mathbb{Z}_2$-even phase. For example, the use of trial
functions (\ref{trial-1}) with $\lambda_v/t=5$ and a $60\times60$
$\k$-mesh results in $\Omega_I=0.02770$ and $\tilde{\Omega}=0.00025$.
We also find that the results are more sensitive to the choice of
$\bf k$-mesh in the $\mathbb{Z}_2$-odd regime.

To summarize the results of this section, we studied the
construction of maximally localized WFs in the $\mathbb{Z}_2$-odd
phase using the Kane-Mele model as an example. We have seen that
our initial guess of Sec.~\ref{Sec.4} about the localization of
WFs in this topological regime is very good, and that the maximal
localization procedure does not greatly reduce the spread.

\section{\label{Sec.6} Hybrid Wannier charge centers and polarization}

In this section we discuss the polarization in $\mathbb{Z}_2$-odd insulators
using the example of the Kane-Mele model,
and see what insights about the topological insulating phase can be
obtained by inspecting this property.

The electronic polarization in a 2D system can be defined either
in terms of the Berry phase\cite{King-Smith-PRB93} 
\begin{equation}
{\bf P}=\frac{|e|}{(2\pi)^2}\mathrm{Im} \sum_{n=1}^{\cal N}\int d{\bf k} 
\langle u_{n{\bf k}} |\nabla_{\bf k}| u_{n{\bf k}}\rangle
\label{polarBerry}
\end{equation}
or via the
summation of Wannier charge centers\cite{Vanderbilt-PRB93}
\begin{equation}
{\bf P}=-\,\frac{|e|}{A}\sum_{n=1}^{\cal N}\bar{\bf r}_n,
\label{polar-wc}
\end{equation}
where $e$ is the electronic charge and $A$ is the area of the unit 
cell. The two definitions are identical and define electronic 
polarization modulo a polarization quantum $|e|\R/A$, $\R$
being a lattice vector. This ambiguity can be understood as a freedom in 
the choice of branch in Eq.~(\ref{polarBerry}) or in the choice of
unit cell in Eq.~(\ref{polar-wc}). The definition via
Wannier charge centers makes the dependence of $\bf P$
on the choice of origin obvious. As described in 
Sec.~{\ref{Sec.3}}, the origin of the Kane-Mele model is chosen
such that atoms are located along the $y$-axis at 
${\bf t}_A=\xi\hat{y}/3$ and ${\bf t}_B=2\xi\hat{y}/3$, where
$\xi=|{\bf a}_1+{\bf a}_2|=a\sqrt{3}$.
Because the Hamiltonian has 3-fold symmetry, we expect the rescaled
polarization $(A/|e|)\bf P$ to lie at the origin, at ${\bf t}_A$,
or at ${\bf t}_B$.  To distinguish between these possibilities
it is sufficient to compute $P_y$, which is well-defined modulo
$|e|/a$.

\subsection{\label{direct} Total polarization}

A direct computation of electronic polarization via 
Eq.~(\ref{polarBerry}) in the $\mathbb{Z}_2$-even phase results in
$P_y = |e|/3a$ mod $|e|/a$, consistent with the fact that
both Wannier centers in Eq.~(\ref{polar-wc}) lie at ${\bf t}_B$
(since $-4|e|\xi/3A=-8|e|/3a=|e|/3a$ mod $|e|/a$.)
In the $\mathbb{Z}_2$-odd phase, on the other
hand, Eqs.~(\ref{polarBerry}) and (\ref{polar-wc})
lead to $P_y=0$ mod $|e|/a$.  Again, this is consistent with
the locations of the WFs.
As indicated in Sec.~\ref{Sec.5}, the Wannier centers $\bar{\bf r}_n$
in this phase lie approximately at ${\bf t}_A$ and ${\bf t}_B$.
More precisely, we find that they are located at
$\bar{\bf r}_1=(1-\delta)\xi\hat{y}/3$ and
$\bar{\bf r}_2=(2+\delta)\xi\hat{y}/3$, where $\delta$ is a small
correction (e.g., $\delta=0.0018$ at $\lambda_v/t=1$).
Thus, the sum of the Wannier centers is just $\xi\hat{y}$,
or zero modulo a lattice vector.

It is interesting to note that, in retrospect, the computation
of the polarization via Eq.~(\ref{polarBerry}) would have given
a strong hint about the appropriate choice of trial functions
in the $\mathbb{Z}_2$-odd insulator.  That is, knowing only that
$P_y=0$, one might have guessed that both WFs should be
centered halfway between ${\bf t}_A$ and ${\bf t}_B$, or both at
the center of the honeycomb ring, or one at ${\bf t}_A$ and the
other at ${\bf t}_B$.  The latter possibility becomes the most
likely when we also take into account that in the $\mathbb{Z}_2$-odd
phase the two WFs cannot form a Kramers pair.

\subsection{\label{hybrid} Hybrid Wannier decomposition}

In order to obtain a deeper understanding
of the origin of the polarization and expose some qualitative
differences in the behavior of its $\k$-dependent decomposition in
$\mathbb{Z}_2$-even and odd phases, it is useful to use a hybrid
representation in which the Wannier transformation is carried out
in one direction only.  As indicated above, we know from symmetry
considerations that we can set $P_x=0$ and characterize the
polarization by $P_y$ mod $\xi|e|/A$.  To compute $P_y$, it is
convenient to choose the BZ to be a rectangle extending over
$k_x\in[0,2\pi/a]$ and $k_y\in[0,4\pi/\xi]$
(corresponding to the region $\zeta$ in Fig.~\ref{ebz}).
We can then define hybrid WFs
\begin{equation}
|nk_x l_y\rangle=\frac{\xi}{4\pi}\int_0^{4\pi/\xi} dk_y \,
   e^{-i k_y l_y} | \bls \rangle
\label{hybridWF}
\end{equation}
in terms of which the usual WFs are
\begin{equation}
|{\bf R}n\rangle = |nl_x l_y\rangle=\frac{a}{2\pi}\int_0^{2\pi/a} dk_x
\, e^{-i k_x l_x} | nk_x l_y \rangle.
\label{WFhybrid}
\end{equation}
The hybrid Wannier centers are defined as
\def\ybar{{\bar{y}_n(k_x)}}
\begin{equation}
\ybar= \langle nk_x0|y|nk_x0\rangle
\label{bynkx}
\end{equation}
and the total electronic polarization is
\begin{equation}
P_y = -\frac{|e|}{\pi\xi}\sum_n \int_0^{2\pi/a} dk_x \, \ybar.
\label{Pdecompy}
\end{equation}
In practice the $k_x$ integral is discretized by a sum over a mesh of
$k_x$ values, and at each $k_x$ the $\ybar$ are calculated by considering
the corresponding string of $\bf k$-points along $k_y$.  In the case
that the gauge has been specified by a particular set of 2D WFs
$|{\bf R}n\rangle$, or, equivalently, by the corresponding
Bloch-like functions $|\tilde{\psi}_{n\bf k}\rangle$, this is
done straightforwardly using the discretized Berry-phase formula
\begin{equation}
\ybar=-\frac{\xi}{4\pi}\,\mathrm{Im}\log\prod_j M^{(j)}_{nn}
\label{berry1D}
\end{equation}
where $M^{(j)}$ is a shorthand for the overlap matrix
$M^{({\bf k}_j,{\bf k}_{j+1})}$ of Eq.~(\ref{M-overlaps}) connecting
$k_y$-points $j$ and $j+1$ along the string.

As was emphasized in Sec.~\ref{Sec.4}, the $\tilde{\psi}_{n{\bf k}}$
carry the information about the gauge choice.  Thus, different
gauge choices -- i.e., different choices of WFs --  will result in
different hybrid WFs and different $\ybar$. However, the sum $\sum_n \ybar$
at a given $k_x$ is gauge-invariant, and as a result $P_y$
of Eq.~(\ref{Pdecompy}) must remain the same in any gauge.

Of special interest is a gauge choice in which, at each $k_x$,
the hybrid WFs $|nk_xl_y\rangle$ are maximally
localized in the $y$ direction.  It was shown in
Ref.~\onlinecite{Marzari-PRB97} that in 1D the Wannier charge centers
can be obtained by a parallel-transport construction using the overlap
matrices $M^{(j)}$.  Specifically, the ``unitary part''
$\widetilde{M}^{(j)}$ of each overlap matrix is obtained
by carrying out the singular-value decomposition $M=V\Sigma
W^\dagger$, where $V$ and $W$ are unitary and $\Sigma$ is
real-positive and diagonal, and then setting
$\widetilde{M}=VW^\dagger$.  This is reasonable because,
for a sufficiently fine mesh spacing, $\Sigma$ is almost
the unit matrix.  Then, the unitary matrix $\Lambda=\prod_{j}
\widetilde{M}^{(j)}$ describes the transport of states along the
string. The eigenvalues $\lambda_n$ of this matrix are all of unit
modulus, and their phases define Wannier centers via\cite{Wu-PRL06}
\begin{equation}
\ybar=-\frac{\xi}{4\pi}\mathrm{Im}\log{\lambda_n}.
\label{wc-p}
\end{equation}
Note that no iterative procedure is needed.
Inserting this equation into Eq.~(\ref{Pdecompy}), one gets a
discretized formula for $P_y$ that is consistent with
Eq.~(\ref{polarBerry}).

\subsection{\label{pol-results} Results}

We illustrate these ideas now for the KM model in its normal and
$\mathbb{Z}_2$-odd phases.  In each case we present results for
$\ybar$ for two choices
of gauge: the maximally-localized one along $\hat{y}$ as discussed
in the previous paragraph, and the one corresponding to the WFs
constructed from the trial functions of
Eq.~(\ref{trial-1}) for the $\mathbb{Z}_2$-even phase or those of
Eq.~(\ref{trial-2}) for the $\mathbb{Z}_2$-odd phase.  In what
follows, we refer to these as the ``maxloc'' and
``WF-based'' gauges respectively.

\begin{figure}
\begin{center}
\includegraphics[width=3.0in, bb= 4 1 600 338]{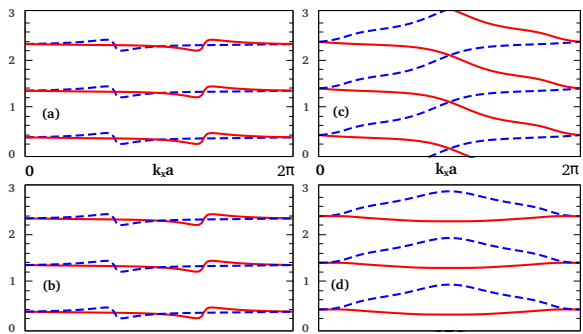}
\end{center}
\caption{(Color online)
Hybrid Wannier centers $\ybar$, in units of $\xi/2$, for
the Kane-Mele model.
$\mathbb{Z}_2$-even phase ($\lambda_v/t=3$):
(a) maxloc gauge; (b) WF gauge of Eq.~(\ref{trial-1}).
$\mathbb{Z}_2$-odd phase ($\lambda_v/t=1$):
(c) maxloc gauge; (d) WF gauge of Eq.~(\ref{trial-2}).
In each case, several periodic images are shown.}
\label{Fig:polar}
\end{figure}

In the ordinary insulating regime, the maxloc and WF-based $\ybar$
curves look very similar to each other.  Fig.~\ref{Fig:polar}(a)
and (b) show the calculated results for the case of $\lambda_v/t=3$,
very close to the transition on the insulating side (recall the critical
value is at $\lambda_v/t=2.93$).  Three of the infinite number of
periodic images along $y$ are shown.  The ``bumps'' in the curves near
the $K$ and $K'$ points in the BZ are the result of the proximity
to the transition; as one goes deeper into the insulating phase, the
curves flatten out and become smooth functions of $k_x$.  The solid and dashed
curves are mirror images of each other; in the maxloc construction
of Fig.~\ref{Fig:polar}(a) this just reflects the time-reversal
invariance of the Hamiltonian, while in Fig.~\ref{Fig:polar}(b)
it follows from the fact that the WFs form a Kramers pair.

When averaged over $k_x$, each curve is found to have a mean $\bar{y}$
value of $2\xi/3$ to numerical precision, or $\xi/6$ modulo
$\xi/2$, consistent with the discussion
in Sec.~\ref{direct}.

The corresponding results for the $\mathbb{Z}_2$-odd phase are
shown in Fig.~\ref{Fig:polar}(c) and (d) for $\lambda_v/t=1$. As
expected, there is again a mirror symmetry visible in the curves
for the maxloc construction in Fig.~\ref{Fig:polar}(c), but the
connectivity of the curves is qualitatively different: in going
from $k_x=0$ to $\pi/a$ we see that the $n$'th solid curve
goes up to cross the $(n+1)$'th dashed curve,
while the $n$'th dashed curve goes down to cross the $(n-1)$'th
solid curve.
This is exactly the kind of behavior that was exhibited in Fig.~3(a)
of Ref.~\onlinecite{Fu-PRB06} as a signal of the
$\mathbb{Z}_2$-odd phase.
Moreover, if we follow the $n$'th dashed curve all the way
across the BZ, we find that it wraps to become the
$(n+1)$'th one when $k_x=2\pi/a$ wraps back to $k_x=0$.
This is precisely the kind of behavior that is characteristic
of a Chern (or quantum anomalous Hall) insulator,\cite{Coh-PRL09}
which implies that we can assign a Chern
number of $+1$ to the
Bloch subspace spanned by the eigenvectors corresponding to the
dashed bands.  However, since we are studying here a system
with time-reversal symmetry, we find also a partner subspace
corresponding to the full curve in Fig.~\ref{Fig:polar}(c)
having Chern number $-1$.  As a result, of course, the overall
occupied space has a total vanishing Chern number, as it must
due to the time-reversal symmetry.  The evaluation of the
polarization $P_y$ through Eq.~(\ref{Pdecompy}) again yields
$P_y=0$ mod $|e|/a$, consistent with the direct calculation
of Sec.~\ref{direct}.

Finally, Fig.~\ref{Fig:polar}(d) shows the $\ybar$ curves for
the same $\mathbb{Z}_2$-odd parameters as in Fig.~\ref{Fig:polar}(c),
but using the WF-based gauge determined by the trial functions of
Eq.~(\ref{trial-2}).  At any given $k_x$, we confirm that
$\bar{y}_1+\bar{y}_2$ is the same in Fig.~\ref{Fig:polar}(d) as in
Fig.~\ref{Fig:polar}(c), and the total polarization is therefore
the same.  However, because the two WFs do not form a Kramers
pair in this case, the dashed and solid curves do not map into each other
under time-reversal symmetry, and there is no degeneracy at
$k_x=\pi/a$.  Moreover, the Chern number of each band is
individually zero, consistent with the fact that each one is
derived from a WF.  The average $\bar{y}$ values for
the solid and dashed curves are
$0.978\xi/3$ and $2.022\xi/3$ mod $\xi/2$, very close to the nominal
locations of the trial functions at ${\bf t}_A$ and ${\bf t}_B$,
respectively.

To recap, in both the $\mathbb{Z}_2$-even and $\mathbb{Z}_2$-odd
cases, we find that the occupied Bloch space can be cast as the
direct sum of two subspaces that map into one another under the
time-reversal operation, corresponding to the solid and dashed
curves of Figs.~\ref{Fig:polar}(a-c).  These subspaces are not
built from Hamiltonian eigenstates, but from suitable
$\bf k$-dependent ${\cal U}(2)$ rotations among the Hamiltonian eigenstates.
In the $\mathbb{Z}_2$-even
case the Chern index of each of these subspaces is separately zero,
so that we can also provide a Wannier
representation for each subspace separately. This is essentially
the case of Fig.~\ref{Fig:polar}(b), and since the spaces form a
time-reversal pair, the WFs form a time-reversal pair as well.
In contrast, for the $\mathbb{Z}_2$-odd phase, the decomposition
into two subspaces that are time-reversal images of each other
necessarily results in subspaces having individual Chern numbers
of $\pm$1, and these are not individually Wannier-representable.
Only by violating the condition that the two spaces be
time-reversal partners, as was done in Fig.~\ref{Fig:polar}(d), can we
decompose the space into two subspaces
having zero Chern indices individually.  By doing so, we can find a Wannier
representation of the entire space, but only on condition that
the two WFs do not form a Kramers pair.

\section{\label{Sec.7} Conclusions}

In this paper we have considered the question of how to construct
a Wannier representation for $\mathbb{Z}_2$-odd topological insulators
in 2D. We have shown that the usual method based on
projection onto trial functions fails because of a topological
obstruction if one imposes the condition that the trial functions
should come in time-reversal pairs.  On the other hand, the projection
method can be made to work if this condition is not imposed,
resulting in WFs that do not transform into one another
under time reversal.

Such a Wannier representation may have some formal disadvantages.
For example, if one writes the Hamiltonian as a matrix in this Wannier
representation, its time-reversal invariance is no longer
transparent, and the presence of other symmetries may become less
obvious as well.  On the other hand, it does satisfy all the usual
properties of a Wannier representation, as for example the ability
to express the electric polarization in terms of the locations of
the Wannier centers, and there is every reason to expect that the
maximally localized WFs are still exponentially localized.
\cite{Brouder-PRL07}

The generalization of our findings to the 3D case should be
relatively straightforward.  Certainly the topological obstruction
to the construction of Kramers-pair WFs remains for both weak and
strong $\mathbb{Z}_2$ topological insulators in 3D.  To see this,
consider in turn each of the six symmetry planes in $\bf k$-space
($k_1=0$, $k_2=0$, $k_3=0$, $k_1=\pi/a$, etc.) on which $H_{\bf
k}$ behaves like a 2D time-reversal invariant system.  For both
weak and strong topological insulators, at least one of these
six planes must have a $\mathbb{Z}_2$-odd 2D invariant.  But if
a gauge exists obeying the time-reversal condition of
Eq.~(\ref{constraint2}) in the 3D $\bf k$-space, then it does so in
particular on the 2D plane, in contradiction with the 2D arguments
about a topological construction.

Thus, the general strategy for constructing WFs for 3D topological
insulators should be very similar to the one presented here in 2D.
Namely, one has to choose pairs of trial functions that do not
transform into one another by time-reversal symmetry, and to do it
in such a way that the projection of these trial functions onto
the Bloch states does not become singular anywhere in the 3D BZ.
While it may be interesting to explore how this might best be done
in practice for real 3D topological insulators, e.g., in the
density-functional context, the choice is likely to
depend sensitively on details of the particular system of interest.
Thus, an investigation of these issues falls beyond the scope of
the present work.

\section{\label{Sec.8} Acknowledgments}

The work was supported by NSF Grant DMR-0549198.

\bibliography{paper}

\end{document}